# Ultracompact high-Q whispering gallery mode microresonator in a non-closed waveguide path

Ziyang Xiong,[1,†] Tong Lin,[1,†,2,*] Liu Li,[3] Hao Deng,[1] Haoran Wang,[1] Yan Fan,[1] Shihua Chen,[4] Junpeng Lu,[1,4] and Zhenhua Ni[1,4]

[1] *School of Electronic Science and Engineering, Southeast University, Nanjing, 210096, China*
[2] *State Key Laboratory of Infrared Physics, Shanghai Institute of Technical Physics, Chinese Academy of Science, Shanghai, 200083, China*
[3] *School of Material Science & Engineering, Southeast University, Nanjing, 210096, China*
[4] *School of Physics, Southeast University, Nanjing, 210096, China*
[†]*Ziyang Xiong and Tong Lin contributed equally to this work.*
*\*lintong@seu.edu.cn*

**Abstract:** Integrated photonic circuits are foundational for versatile applications, where high-performance traveling-wave optical resonators are critical. Conventional whispering-gallery mode microresonators (WGMRs) confine light in closed-loop waveguide paths, thus inevitably occupy large footprints. Here, we report an ultracompact high loaded Q silicon photonic WGMR in an open curved path instead. By leveraging spatial mode multiplexing, low-loss mode converter-based photonic routers enable reentrant photon recycling in a single non-closed waveguide. The fabricated device achieves a measured loaded Q-factor of $1.78 \times 10^5$ at 1554.3 nm with a 1.05 nm free spectral range in a ultracompact footprint of 0.00137 mm$^2$–6× smaller than standard WGMRs while delivering 100× higher Q-factor than photonic crystal counterparts. This work pioneers dense integration of high-performance WGMR arrays through open-path mode recirculation.

## 1. Introduction

In recent years, silicon photonics, owing to its exceptional compatibility with standard complementary metal-oxide-semiconductor processes and its high integration density, has become a pivotal research platform in the field of photonic integrated circuits [1,2]. Travelling-wave optical resonators on the silicon-on-insulator (SOI) platform leverage the high refractive index contrast at the Si/SiO$_2$ interface to deliver high Q-factors and large free spectral range (FSR) in compact designs. With expanding applications spanning optical interconnects [3], wavelength add/drop filters [4–6], optical sensors [7], hybrid external cavity lasers [8–11], quantum computing [12], and microwave photonics [13], demand intensifies for resonators that simultaneously provide higher Q-factors, moderate FSR, and more compact resonators for comprehensive system-level integration [14].

Fundamentally, WGMRs rely on self-constructive interference of transverse modes within a closed waveguide path, traditionally requiring multiple bends for 360° light routing. Achieving high Q-factors demands extended optical paths to suppress bending loss [15], inevitably increasing resonator footprints. This larger size inversely reduces the FSR (i.e., FSR ∝ 1/perimeter). While early WGMRs adopted circular geometries [16], the isoperimetric inequality dictates that non-circular shapes (e.g., ellipses) with identical perimeters enclose smaller areas. To reconcile compactness with performance, multimode free-form waveguides optimize bending radii [17–19], yet their enclosed areas remain constrained by 180° bend lateral offsets (twice the bending radius). To miniaturize WGMR, traveling wave wire waveguide resonators employ mode-converting sidewall Bragg grating reflectors [20] or photonic crystal nanobeam reflectors [21,22], compressing the traditional closed-curve area to single-waveguide dimensions via modal multiplexing. Although eliminating bends enables extreme footprint reduction, these Fabry-Perot-like cavities suffer from reflector-induced excess losses beyond propagation loss [23], yielding significantly degraded Q-factors. Notably, these subwavelength elements demand stringent nanofabrication—incompatible with scalable CMOS production. Thus, achieving high-Q factors in ultracompact silicon photonic WGMRs remains a critical challenge.

In this study, we propose an ultracompact high-Q WGMR operating in an open curve waveguide. Utilizing the spatial mode multiplexing technique, low-loss mode converters establish a closed photonic loop in a linear topology as shown in Fig. 1(a), corresponding to a footprint of merely 0.00137 mm². Our design integrates an asymmetric directional coupler (ADC), broadband adiabatic mode converter (AMC) reflectors, and multimode waveguide to collectively suppress scattering and higher-order mode losses. Compared to sidewall Bragg grating or photonic crystal reflectors, this AMC reflector design significantly reduces excess losses: the simulated mode conversion efficiency reaches up to 99.98% (minimum reflectivity>97.9%) within the 1500–1600 nm wavelength range. Experimental measurements reveal that the device achieves the highest loaded Q-factor of $1.78 \times 10^5$ at 1554.3 nm and is thermo-optically tuned over one FSR (1.051 nm) with 1.31 pm/mW efficiency. Furthermore, the minimum feature size of the device is 135 nm, compatible with standard DUV lithography. Combined with high-Q operation, broadband performance, extreme compactness, and fabrication resilience simultaneously, this platform holds promise in on-chip photonic signal routing and processing using dense WGMR arrays.

## 2. Design and principle

We design an ultra-compact high-Q WGMR within an open curve waveguide by leveraging the mode multiplexing technique [24–26]. As illustrated in Fig. 1(a), the device integrates a single-mode bus waveguide and two AMC reflectors interconnected by a multi-mode waveguide; the two adjacent waveguides form an ADC. Crucially, three mode converters establish a unidirectional optical loop by interconverting between $TE_0$ and $TE_1$ modes, as shown in the spatial mode domain (Fig. 1(b)). The process initiates when the ADC evanescently injects a small portion of input $TE_0$ mode into the $TE_1$ mode of the resonant path based on the phase match. This $TE_1$ mode propagates rightward to the AMC reflector, where it is converted to $TE_0$ through adiabatic eigenmode transformation and redirected via a curved waveguide toward the left port. Nobly, the ADC functions as an all-pass filter for $TE_0$, allowing straight-through transmission to the left AMC. There, $TE_0$ is reconverted to $TE_1$ with high efficiency. At the end of the first roundtrip, the ADC couples a small port of $TE_1$ back to the output $TE_0$ mode, with most of $TE_1$ repeating this process inside the optical loop. In this way, the interference cycle is completed. This unidirectional recycling path mimics a microring resonator within an open curve configuration, demonstrating pronounced compactness as compared to the closed curve conventional whispering-gallery mode resonant (Fig. 1(c)). Meanwhile, the high Q-factor, broad bandwidth, and fabrication simplicity are simultaneously attained through optimized mode converters, overcoming classical footprint trade-offs in conventional microring resonator design.

We employ a pair of AMC reflectors with the mode conversion-based reflectivity up to 99.98% to suppress the excess losses. Unlike the interferometric directional coupler for spatial mode conversion, these adiabatic directional couplers [27,28], eliminate stringent phase-matching requirements and precise geometric control. Figure 2(a) details the AMC schematic with key parameters. We carried out three-dimensional finite-difference time-domain simulations to evaluate its performance. The simulated mode conversion efficiencies and transmissions for $TE_0$ and $TE_1$ across 1500–1600 nm range are shown in Fig. 2(b). For $TE_1$ input, the minimum conversion loss is 0.013 dB (i.e., conversion efficiency 99.98%) and the maximum conversion loss is 0.128 dB (conversion efficiency 97.9%), with the intramode extinction ratio (ER)< –17 dB.

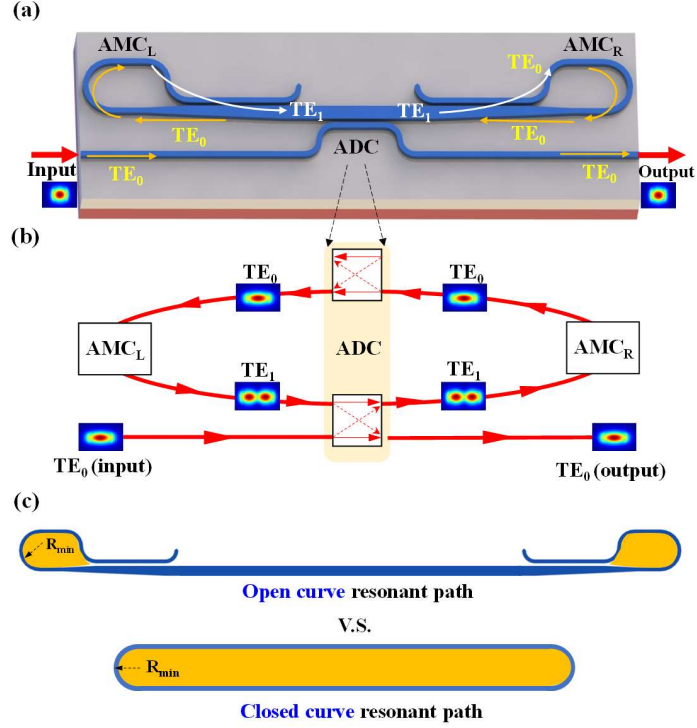

Fig. 1. (a) A schematic diagram of the proposed WGMR structure. It manifests the open curve light-recycling path. (b) The operational principle of the proposed WGMR structure with three mode converters inside. It shows the light propagation path in the closed curve inside the spatial mode domain. (c) The comparison between the ultracompact proposed open curve resonant and a conventional closed curve resonator. The yellow-colored areas highlight the occupied footprint.

For $TE_0$ input, the transmission loss ranges from -0.014 dB (transmission 99.97%) to -0.012 dB (transmission 99.98%), achieving an intramode ER below –40 dB. The time-integrated mode field distributions of the AMC for $TE_1$ and $TE_0$ inputs at a wavelength of 1550 nm (Fig. 2(c)) visually confirm highly efficient mode conversion and transmission for two modes respectively. Furthermore, the 9.44-μm-radius ($R_2$) curved waveguide connecting drop and through ports introduces negligible bending loss. Combined with broadband reflectivity (peak 99.98%) and a 135-nm coupling gap ($gap_1$) fabricable by standard DUV lithography, this reflector design delivers unprecedented performance accessibility. This coupling gap can be further enlarged via increased coupling length ($L_1$) without sacrificing the device compactness as the coupling length is also part of the cavity.

We implement a low-loss mode-converting directional coupler with ultrahigh mode-selectivity and mediated cross-coupling to achieve high Q-factors. Figure 2(d) shows the ADC's structural parameters with two distinct optical paths: the $TE_0$ mode exhibits near-unity transmission (>99.99%, intramode ER<-55 dB) for minimized propagation loss, while the $TE_1$ mode undergoes controlled partial conversion to $TE_0$ (<3% cross-coupling at 1550 nm) while retaining >97% transmission to balance with the cavity intrinsic loss. Figure 2(e) presents simulated transmission spectra for two modes at two ports respectively; they confirm these characteristics with intermodal crosstalk below -39 dB and insertion losses under -0.03 dB for both modes across a 100-nm bandwidth, collectively ensuring the broadband low-loss operation essential for the proposed high-Q WGMR. As a qualitative demonstration of unidirectional propagation, the time-integrated mode field profiles through the ADC for two orthogonal mode inputs at a wavelength of 1550 nm are shown in Fig. 2(f), emphasizing high efficiency transmissions.

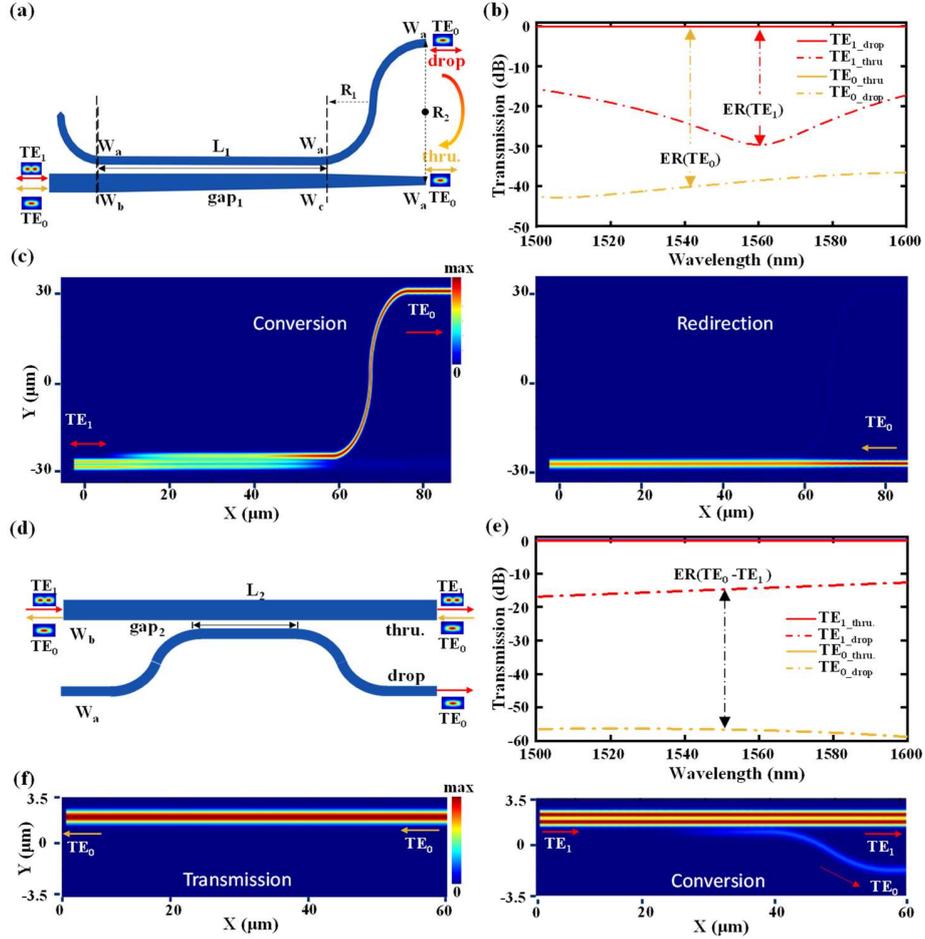

Fig. 2. (a) Schematic illustration of the AMC ($W_a$=0.495 μm, $W_b$=1.08 μm, $W_c$=0.975 μm, $gap_1$=0.135 nm, $L_1$=53 μm, $R_1$=9 μm, $R_2$≈9.44 μm). (b) Simulated mode-conversion efficiency of the AMC over the 1500–1600 nm spectral range. (c) Mode-field distributions of the AMC for $TE_1$ and $TE_0$ inputs at a wavelength of 1565 nm. (d) Schematic illustration of the ADC ($W_a$=0.495 μm, $W_b$=1.08 μm, $gap_2$=0.275 μm, $L_2$=14 μm). (e) Simulated transmission spectra of the ADC over the 1500–1600 nm spectral range. (f) Mode-field distributions of the ADC for $TE_0$ and $TE_1$ input inside one light round trip at a wavelength of 1550 nm.

## 3. Experiment results

We characterize optical properties of the WGMR using the experimental setup schematically illustrated in Fig.3. Fabricated on a 220-nm SOI platform multi-project-wafer service (Applied Nanotools Inc), the device under test (DUT) occupies a compact footprint of 0.00137 $mm^2$. Light from a narrow-linewidth tunable laser source (Santec TLS-570) enters the chip via a grating coupler, with polarization adjusted by a polarization controller. The output signal from the chip is emitted by a second grating coupler, which was coupled by another optical fiber and detected subsequently in a photodetector module (Santec MPM-210). All measured transmission spectra are normalized with respect to those obtained from a U-shaped reference waveguide in proximity. The left inset in the figure shows a photograph of the chip, while the right one displays the experimental optical coupling setup with fibers aligned with grating couplers at optimized angles. In addition, DC electrical probes were landed onto the DUT's metal pads to apply electrical powers to the chip for effective cavity length tuning.

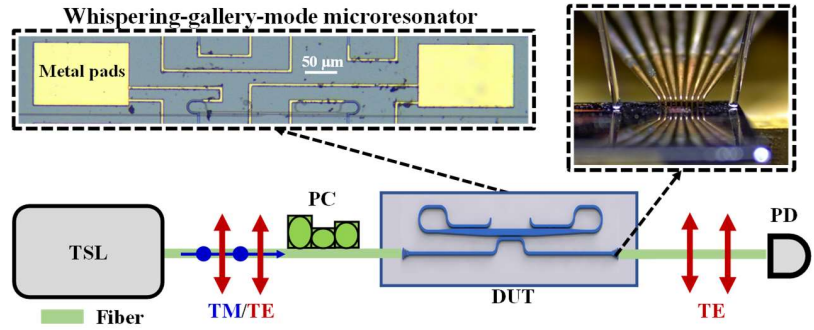

Fig. 3. The schematic diagram of the experimental setup for device characterizations (TSL: tunable laser source, PC: polarization controller; DUT: device under test; PD: photodetector). The insets are the microscope images of the device (top view) and the fiber coupling section (side view).

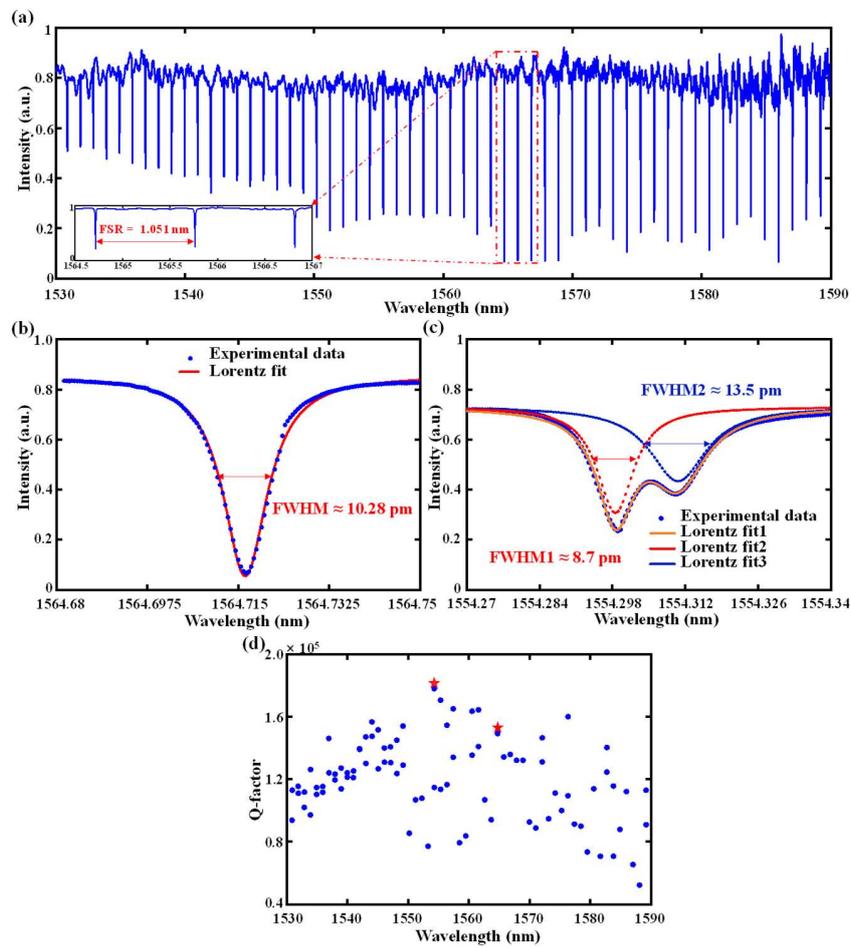

Fig. 4. (a) Experimental transmission spectrum of the WGMR for the optical input power of 0 dBm. The inset presents an enlarged view of a selected spectral window, highlighting its FSR. (b) Lorentzian fit of the single resonance peak at 1564.715 nm with its FWHM indicated. (c) Lorentzian fit of the mode-splitting resonance at 1554.3 nm, with their FWHMs indicated respectively. (d) Loaded Q-factors of resonances extracted from Lorentzian fit for the wavelength from 1530 nm to 1590 nm.

We measured the highest loaded Q-factor exceeding $1.78 \times 10^5$ at 1554.3 nm with a FSR of about 1.051 nm in the fabricated device. The experimental transmission spectrum of the WGMR from 1530 nm to 1590 nm is shown in Fig. 4(a). The inset highlights two adjacent resonances near 1565 nm, confirming the WGMR's FSR. Resonance extinction ratio variations stem primarily from wavelength-dependent ADC efficiency—despite near-constant AMC reflectivity—and can be mitigated via dispersion-flattened AMC designs. Lorentz fitting (Fig. 4(b)-(c)) quantifies two representative resonances: a single dip at 1564.715 nm yields a full width at half maximum (FWHM) of 10.3 pm (loaded Q-factor $\approx 1.52 \times 10^5$ ($Q_L = \lambda_{res}/\text{FWHM}$), finesse F = 102.2), while a split-mode resonance at 1554.3 nm (attributed to fabrication-induced backscattering [29] achieves a higher loaded Q-factor of $1.78 \times 10^5$ with 8.7-pm FWHM and finesse F = 120.8 extracted from the double dip Lorentz fitting results. Here, finesse F quantifies the cavity's spectral selectivity and energy-storage capability. We further calculate the loaded Q-factors for all the resonances within the 1530–1590 nm wavelength range and summarize them in Fig. 4(d). Statistical analysis of all resonances confirms consistently high loaded Q-factors exceeding $5 \times 10^4$ across the band, reaching their maximum near 1554.3 nm.

We experimentally demonstrate resonance-linewidth broadening induced by thermo-optic nonlinearity—another key indicator of a high Q factor. Figure 5 presents transmission spectra of the WGMR near 1565.63 nm under laser output powers of −5 dBm, 4 dBm, and 13 dBm. At −5 dBm, the resonance exhibits a symmetric Lorentzian dip, indicating operation in the linear regime. As the input power increases to 4 dBm and 13 dBm, linear absorption within the cavity produces local heating; because silicon has a positive thermo-optic coefficient, the refractive index rises with temperature and the effective optical path length increases, shifting the resonance toward longer wavelengths. Meanwhile, the thermal-diffusion response lags the scan, leading to thermal dragging of the resonance, manifested as mild lineshape asymmetry and a red shift of the resonance. These results demonstrate pronounced nonlinear phenomena occurring at comparatively low pump powers in this high-Q WGMR, providing experimental insights into dynamic bistability and nonlinear behavior in integrated photonics [30,31,36].

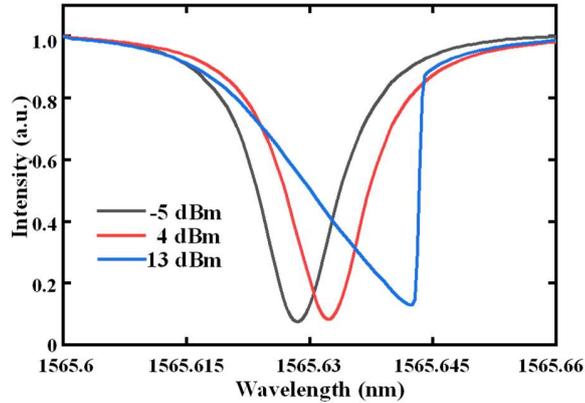

Fig. 5. Nonlinear response of the resonance peak under varying pump powers emitted from the laser source.

The WGMR exhibits thermal tuning exceeding one FSR with applied power below 81 mW. As shown in Fig. 6(a), the resonance peak wavelength red-shifts from 1566.64 nm to 1568.08 nm when the applied power is increased to 104.29 mW, yielding a tuning span of around 1.44 nm. Figure 6(b) quantifies this linear response, revealing a tuning efficiency of 13.1 pm/mW with strong linear correlation ($R^2$=0.9982). Critically, full $2\pi$ phase shifts require less than 81 mW. The observed fluctuations in extinction ratio originate from thermal crosstalk to the AMCs, which perturbs their optical mode-conversion efficiency and thereby changes the Q-factors of the resonant peaks. These discrepancies can be compensated with another two microheaters that regulate the AMCs cohesively.

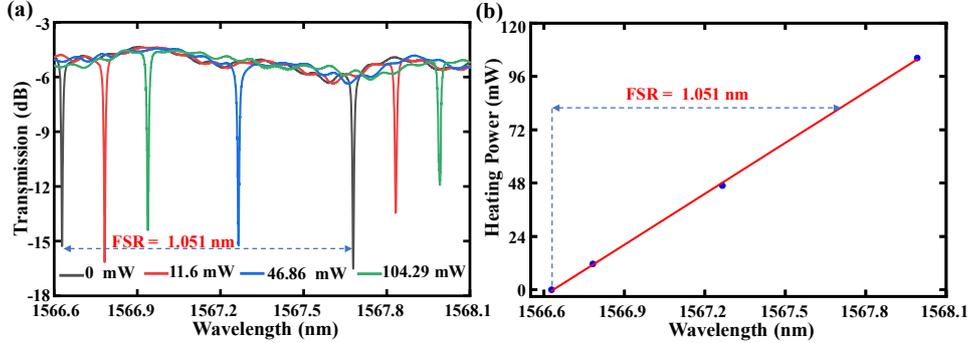

Fig. 6. (a) Transmission spectra of the WGMR measured under different applied electrical powers. (b) Resonance peak wavelength shifts linearly as a function of the applied electrical power.

Figure 7 visually underscores the superior compactness of our open-path resonator design when implemented in a wavelength division multiplexing add-drop filter array. The upper panel illustrates an array constructed from a cascaded series of our open-path resonators ($D_1$, $D_2$, … $D_n$), which efficiently extract individual wavelengths ($\lambda_1$, $\lambda_2$, … $\lambda_n$) from the main bus waveguide. As the output directions of the bus waveguide and the drop waveguides are the same, high scalability is ensured. In stark contrast, the conventional wavelength division multiplexing architecture employing microring resonators (MRRs) is illustrated in the lower panel. Owing to the traveling-wave nature of MRRs, the dropped light exits in a direction that is 180º reversed relative to the output of the bus waveguide. Consequently, this design necessitates not only a discrete array of physically isolated MRRs but also a series of complementary U-bend waveguides to achieve the same multiplexing functionality. This side-by-side comparison highlights a fundamental advantage: our architecture eliminates the large interstitial spaces inherent to systems of isolated MRRs and U-bends, enabling a drastically more dense and scalable integration footprint for on-chip spectral filtering systems. This exceptional compactness is a direct consequence of the open-path design, which avoids the traditional area penalty of closed-loop resonators and is a key enabling feature for future high-channel-count communication and signal processing applications.

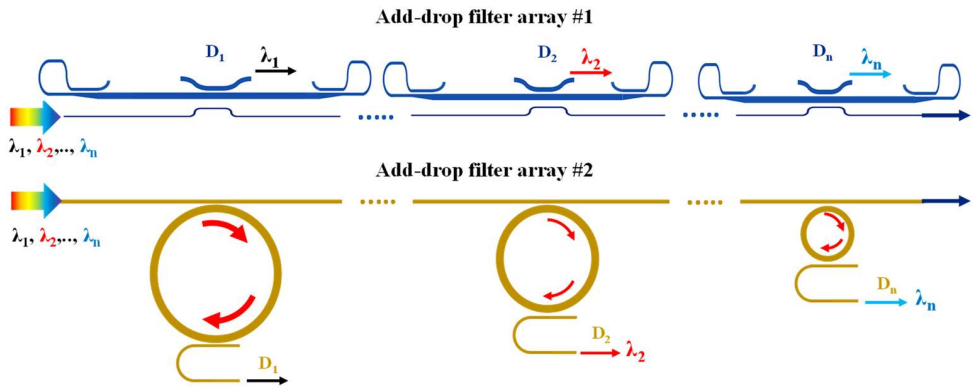

Fig. 7. Illustration of compactness advantage of an add-drop filter array using our open-path resonator design. The lower panel shows the conventional design using multiple microring resonators.

Table 1 summarizes the performance of state-of-the-art near-infrared chip-scale traveling-wave microresonators on SOI platforms. Photonic crystal nanobeam cavities achieve the smallest footprint by eliminating bending waveguides. However, mode conversion occurs within their extremely small 1-D photonic crystals, which require minimal features of 40 nm [6] and thus stringent nanofabrication techniques like electron-beam lithography. Currently, only one experimental work is available to the best of our knowledge, reporting a Q-factor of just 1700—over two orders of magnitude lower than our design. Similarly, sidewall-corrugated Bragg grating resonators [20] exhibit comparably low Q-factors. Conventional MRRs achieve higher Q-factors, peaking at $2\times10^6$ using Euler bends ($R_{eff} \approx 29$ μm) and multimode waveguides [18]. Yet, their closed-curve configuration intrinsically limits miniaturization, resulting in footprints six times larger than ours. Furthermore, our approach allows adopting similar techniques to minimize bending loss within the AMC reflectors, thereby boosting Q-factor in the same way. Collectively, our design offers significant advantages in optical performance, compactness, and relaxed fabrication requirements. This enhances tolerance to process fluctuations and improves consistency and manufacturability for arrayed devices.

**Table 1. Performance comparison of on-chip travelling wave cavities working at infrared wavelengths.**

| Reference | Resonator type | Q-factor | FSR [nm] | Size | Platform | Exp. |
| --- | --- | --- | --- | --- | --- | --- |
| [21] | Photonic crystals nanobeam mode converter | 2250 | 28.5 | 28 μm$^2$ | SOI | No |
| [6] | Photonic crystals nanobeam mode converter | 1700 | 36 | 34.5 μm$^2$ | SOI | Yes |
| [32] | Photonic crystals nanobeam mode converter | 2755 | / | / | SOI | No |
| [33] | Photonic crystals nanobeam mode converter | 4310 | / | / | SOI | No |
| [20] | Width-corrugated Bragg grating mode converter | ~3100 | ~5 | 130.3 μm$^2$ | SOI | Yes |
| [34] | MRR | $1.16\times10^6$ | 0.325 | 0.05 mm$^2$ | SOI | Yes |
| [18] | MRR | $1.3\times10^6$ | 0.9 | ~0.009 mm$^2$ | SOI | Yes |
| [35] | MRR | $2.8\times10^5$ | 3.25 | 0.00785 mm$^2$ | SOI | Yes |
| this work | Adiabatic mode converter | $1.78\times10^5$ | 1.051 | 0.00137 mm$^2$ | SOI | Yes |

## 4. Conclusions

In summary, we demonstrate a traveling-wave whispering-gallery mode resonator operating via open-path mode recirculation enabled by spatial mode multiplexing. By implementing low-loss mode converters as photonic routers, this design establishes a closed optical loop within a linear waveguide topology, eliminating same-mode interference while achieving an ultracompact footprint of 0.00137 mm$^2$. Simulations confirm >99.98% mode-conversion reflectivity and >99.99% mode-sensitive transmission across 1500–1600 nm. Experimentally, the device attains a loaded Q-factor of $1.78\times10^5$ at 1554.3 nm (FSR≈1.051 nm), with full FSR thermo-optic tuning at 13.1 pm/mW efficiency. By suppressing excess losses through open-path recirculation, this architecture offers flexible layout and high yield—enabling scalable WGMR arrays for on-chip linear/nonlinear photonic signal processing.

**Funding.** This work was supported by the National Natural Science Foundation of China under Grant 62105061 and 12374301, in part by the Open Fund of State Key Laboratory of Infrared Physics under Grant SITP-SKLIP-YB-2025-08, and by the National Key Research and Development Program of China 2024YFA1210500.

**Disclosures.** The authors declare no conflicts of interest.

**Data availability.** Data availability. Data underlying the results presented in this paper are not publicly available at this time but may be obtained from the authors upon reasonable request.